\documentclass[numberedappendix,appendixfloats,10pt]{emulateapj}
\usepackage{bm}
\usepackage{pslatex}
\usepackage{graphicx}
\usepackage{natbib}
\usepackage{capt-of}
\begin{document}
\title{Limits on Alpha Particle Temperature Anisotropy and Differential Flow from Kinetic Instabilities: Solar Wind Observations}

\author{Sofiane Bourouaine\altaffilmark{1}, Daniel Verscharen\altaffilmark{1}, Benjamin D. G. Chandran\altaffilmark{1}, Bennett A. Maruca\altaffilmark{2} \& Justin C. Kasper \altaffilmark{3}}
\altaffiltext{1}{ Space Science Center, University of New Hampshire, Durham, NH 03824, USA.}
\altaffiltext{2}{Space Science Laboratory, University of California, Berkeley, CA 94720, USA.}
\altaffiltext{3}{Harvard-Smithsonian Center for Astrophysics, Cambridge, MA 02138, USA}

\altaffiltext{1}{email: s.bourouaine@unh.edu}

%

\begin{abstract}
Previous studies have shown that the observed temperature anisotropies
of protons and alpha particles in the solar wind are constrained by
theoretical thresholds for pressure-anisotropy-driven instabilities
such as the Alfv\'en/ion-cyclotron (A/IC) and
fast-magnetosonic/whistler (FM/W) instabilities.  In this letter, we
use a long period of in-situ measurements provided by the {\em Wind}
spacecraft's Faraday cups to investigate the combined constraint on
the alpha-proton differential flow velocity and the alpha-particle
temperature anisotropy due to A/IC and FM/W instabilities.  We show
that the majority of the data are constrained to lie within the region
of parameter space in which A/IC and FM/W waves are either stable or
have extremely low growth rates. In the minority of observed cases in
which the growth rate of the A/IC (FM/W) instability is comparatively
large,  we find relatively higher values of $T_{\perp\alpha}/T_{\perp p}$ ($T_{\parallel\alpha}/T_{\parallel p}$)
when alpha-proton differential flow velocity is small, where $T_{\perp\alpha}$ and $T_{\perp p}$ ($T_{\parallel\alpha}$ and $T_{\parallel p}$)
are the perpendicular (parallel) temperatures of alpha particles and protons.
We conjecture that this observed feature might arise from preferential alpha-particle 
heating which can drive the alpha particles beyond the instability thresholds.
\end{abstract}

\maketitle



\vspace{0.2cm} 
\section{Introduction}
\vspace{0.2cm}

 In situ spacecraft measurements indicate that
solar-wind plasma deviates significantly from local thermodynamic
equilibrium (LTE). Ions exhibit distinct non-thermal
kinetic features, such as proton core temperature anisotropy, proton
beams, and the preferential heating and acceleration (with respect to
the protons) of alpha particles and minor ions
\citep{marsch06}. All these non-thermal features can be a source of
kinetic instabilities, such as the Alfv\'en/ion-cyclotron (A/IC),
mirror-mode, fast-magnetosonic/whistler (FM/W), and oblique firehose (FH) instabilities.

During the transit of the ions from the Sun to a heliospheric
distance~$r$ of 1 AU, the adiabatic expansion of the solar wind tends
to drive a temperature anisotropy of the form $T_{\parallel}>T_\perp$,
where $\parallel$ and $\perp$ refer to the directions parallel and
perpendicular to the local magnetic field~\citep{chew56}.  On the other
hand, ions can be imparted with the opposite sense of temperature
anisotropy by heating from either the dissipation of low-frequency
turbulence \citep{dmitruk04,parashar09,bourouaine08,chandran10a} or
resonant cyclotron interactions with high-frequency~A/IC
waves~\citep{isenberg01,marsch01,hollweg02,gary06}.  {\em Wind}
measurements at 1 AU reveal that the temperature anisotropy of protons
and alpha particles can be a source for A/IC, mirror-mode, FM/W, and oblique FH
instabilities \citep{kasper02,hellinger06,bale09,maruca12}.  Also, {\em
  Helios} measurements near 0.3 AU show that the velocity distribution
functions of the proton core in fast solar wind regions are close to
marginal stability for the A/IC instability \citep{bourouaine10}.

In situ measurements in the inner heliosphere indicate that alpha
particles can be accelerated up to the local Alfv\'en speed in the
proton frame \citep{marsch82,
  neugebauer94,bourouaine11a,bourouaine11b}. However, the differential
speed between alpha particles and protons rarely exceeds the local
Alfv\'en speed, because super-Alfv\'enic alpha-particle beams lead to
the excitation of A/IC and FM/W waves
\citep{li00,gary00,verscharen13b}, and the amplified waves can
decelerate the alpha particles \citep{kaghashvili04,lu09}. 

 Collisions in the solar wind tend to equilibrate the plasma to a state far below the instability thresholds. Nevertheless, 
the collisionally regulated plasma can still excite waves if it approaches a threshold for instabilities with sufficiently high growth rates.

Some previous studies have focused on instabilities driven by either
an alpha-particle temperature anisotropy $R_\alpha = T_{\perp
  \alpha}/T_{\parallel \alpha} \neq 1$ or a non-zero average
alpha-particle velocity $U_\alpha$ in the proton rest frame. However,
other studies have shown that temperature anisotropy modifies the
$U_\alpha$ threholds of the A/IC and FM/W instabilities, while
differential flow modifies the~$R_\alpha$ thresholds of these
instabilities \citep{gary93,araneda02,gary03, hellinger03,
  verscharen13a}. Our goal in this Letter is thus to treat $U_\alpha$
and $R_\alpha$ on an equal footing and, by analyzing data from the
{\em Wind} spacecraft, to determine whether the linear A/IC
and FM/W instabilities provide a good explanation for the limits on
$U_\alpha$ and $R_{\alpha}$ that are observed in the solar wind.

\vspace{0.2cm} 
\section{Observations and results}
\vspace{0.2cm}

The measurements of ion parameters used in this study were derived from in situ data from the {\em Wind} spacecraft's Faraday cups \citep{ogilvie95}.  This instrument produces an ion spectrum (i.e., a distribution of ion speeds projected along various axes) about once every ninety seconds.  The bulk parameters (e.g., density, flow velocity, and temperature) of the protons and alpha particles can be deduced from each spectrum by fitting a model velocity distribution function (VDF) for each species \citep{phdt:kasper,kasper06}.  Perpendicular and parallel temperature components can be separated using measurements of the local magnetic field, which are available from \textit{Wind}'s Magnetic Field Investigation \citep{lepping95}.

For this study, we used the dataset of ion parameters produced by \citet[][Chapter 4]{phdt:maruca}, who processed nearly $4.8$-million {\em Wind} ion spectra (i.e., all spectra from the spacecraft's launch in late-1994 through mid-2010) with a fully-revised fitting code.  These revisions dramatically improved the code's analysis of temperature anisotropy and differential flow (especially during periods of significant fluctuations in the background magnetic field) \citep{maruca13}.  Nevertheless, only about $2.1$-million of the spectra processed were included in the final dataset due to two sets of selection requirements.  First, a spectrum needed to have been measured at a time when {\em Wind} was well outside the Earth's bow shock (i.e., actually in the solar wind).  The spacecraft, especially during the early part of its mission, spent significant amounts of time exploring the Earth's magnetosphere.  Second, the fit results had to be of high quality as gauged by reduced-$\chi^2$, uncertainty in the fit parameters, and other metrics.  The most frequent cause for this second criterion not being met was low alpha-particle signal (from, e.g., low densities or high temperatures).

To study the instabilities resulting from a combination of relative drift and temperature anisotropy of alpha particles we restrict our data analysis to solar-wind intervals in which $3 \le T_\alpha/T_{\rm p} \le 5$. The selected interval of $T_\alpha/T_{\rm p}$ represents the typical range of variation of the ratio alpha-to-proton temperature in weakly collisional solar wind streams  \citep{kasper08}. Also, our selection is consistent with the theoretical value of $3\lesssim T_\alpha/T_{\rm p}\lesssim5$ used below to determine the threshold values for the drift-anisotropy instabilities.

\begin{figure}
 \includegraphics[width=13pc,height=26pc]{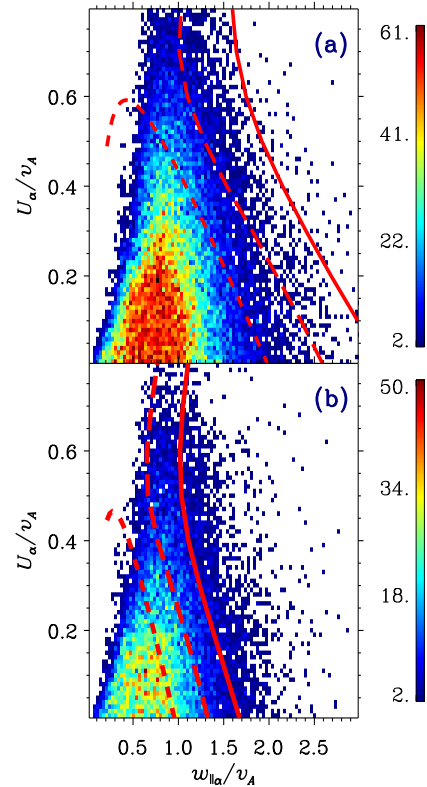}
\caption{Distribution of data in the $U_\alpha - w_{\parallel \alpha}$
  plane, where $U_\alpha$ is the alpha-proton drift speed and
  $w_{\parallel \alpha}$ is the parallel alpha-particle thermal
  speed. The number of measurements in each bin is shown in the color
  bars on the right.  The top panel is the subset of the data in which
  $1.1 < T_{\perp \alpha}/T_{\parallel \alpha} < 1.3$, and the bottom
  panel is the subset of the data in which $1.3 < T_{\perp
    \alpha}/T_{\parallel \alpha} < 1.5$.  The short-dashed lines are
  plots of the A/IC instability threshold (Equation~(\ref{eq:Umin}))
  with $T_{\perp \alpha}/T_{\parallel \alpha} = 1.2$ (top panel) and
  $T_{\perp \alpha}/T_{\parallel \alpha} = 1.4$ (bottom panel).  The
  long-dash (solid) line corresponds to parameter combinations for
  which the maximum A/IC growth rate is $\gamma = 10^{-3}\Omega_{\rm p}$
  ($\gamma =3\times 10^{-3}\Omega_{\rm p}$), where again $T_{\perp
    \alpha}/T_{\parallel \alpha} = 1.2$ in the top panel and $T_{\perp
    \alpha}/T_{\parallel \alpha} = 1.4$ in the bottom
  panel.} \label{fig.1}
\end{figure}

Using Kennel \& Wong's (\citeyear{kennel67}) expression for the growth rate $\gamma$ of
weakly growing waves, \cite{verscharen13a} derived approximate
analytic expressions for the instability thresholds of A/IC and FM/W
waves taking into account both the alpha-proton drift and
alpha-particle temperature anisotropy. For this calculation, they
assumed that the wavevector~$\bf{k}$ is parallel to the background
magnetic field~$\bf{B}_0$ and took the alpha-particles (protons) to
have a bi-Maxwellian (Maxwellian) distribution. We note at this point that some authors refer to the parallel FM/W instability as the \emph{parallel firehose instability}. \cite{verscharen13a} validated their analytic results
by comparing them to numerical solutions of the hot-plasma dispersion
relation. For the parameters we consider in this Letter the minimum
value of $U_\alpha$ needed to excite the A/IC instability is given by
(see \cite{verscharen13a} for further details)

\begin{equation} 
U_{\rm A/IC}=v_{\mathrm A}-\sigma \left(R_\alpha-1\right)w_{\parallel\alpha}-\frac{v_{\mathrm A}^2}{4\sigma w_{\parallel\alpha}R_\alpha},
\label{eq:Umin} 
\end{equation} 
where $w_{\parallel j}= (2 k_B T_{\parallel j}/m_j)^{1/2}$ is the parallel
thermal speed and $m_j$ the mass per particle of species $j$. The minimum value of $U_\alpha$
needed to excite the parallel FM/W instability is
\begin{equation} 
U_{\rm FM/W} = v_{\mathrm A}-\sigma \left(1-R_\alpha \right)w_{\parallel\alpha}+\frac{v_{\mathrm A}^2}{4\sigma w_{\parallel\alpha}R_\alpha}.
\label{eq:UFMW} 
\end{equation} 
The value of the dimensionless quantity~$\sigma$ in these equations
depends very weakly upon the alpha-to-proton density
ratio~$n_\alpha/n_{\rm p}$ and the exact definition of the instability
threshold. In this Letter, we use the values of~$\sigma$ for which
Equations (\ref{eq:Umin}) and (\ref{eq:UFMW}) correspond to growth
rates of $10^{-4} \Omega_{\rm p}$ (where $\Omega_{\rm p}$ is the
proton cyclotron frequency) in a plasma with~$n_\alpha/n_{\rm p} =
0.05$. These values are $\sigma=2.4$ in Equation~(\ref{eq:Umin}) and
$\sigma=2.1$ in Equation~(\ref{eq:UFMW})~\citep{verscharen13a}. In addition to these approximate analytic instability thresholds, 
we use numerical solutions of the hot-plasma dispersion relation to find contours in
different parameter planes (e.g., the $U_\alpha - w_{\parallel
  \alpha}$ plane) corresponding to various values of the maximum A/IC
or FM/W growth rate.  To solve the linear dispersion relation we
  used the following parameters: $n_\alpha/n_{\rm p}=0.05$, $T_{\rm e}=T_{\rm p}$, $R_{\rm p}=1$, $T_{\parallel\alpha}=4T_{\rm p}$, and $v_A/c=10^{-4}$, where $c$ is the speed of light. We plot
some of these contours in the figures below.
As shown by \cite{verscharen13a}, Equations~(\ref{eq:Umin}) and (\ref{eq:UFMW}) 
correspond closely to the numerical contours with $\gamma=10^{-4} \Omega_{\rm p}$, except for the portion of the analytic curve for $U_{\rm A/IC}$ at small
$w_{\parallel \alpha}/v_{\rm A}$ in Figure~\ref{fig.1}  where $U_{\rm A/IC}$ decreases
as $w_{\parallel \alpha}/v_{\rm A}$ decreases, which is not reproduced in the numerical solutions.

In Figure~\ref{fig.1}, we compare the theoretical instability
threshold of the A/IC wave with the subsets of the {\em Wind}
measurements in which $1.1 < T_{\perp \alpha}/T_{\parallel \alpha} <
1.3$ (top panel) and $1.3 < T_{\perp \alpha}/T_{\parallel \alpha} <
1.5$ (bottom panel). We also plot curves corresponding to maximum A/IC growth rates of $10^{-3} \Omega_{\rm p}$ and $3\times 10^{-3} \Omega_{\rm p}$.
When $1.1 < T_{\perp \alpha}/T_{\parallel \alpha} < 1.3$, the large
majority of the data is constrained by the analytic threshold in
Equation~(\ref{eq:Umin}), corresponding to a maximum A/IC growth rate
of $10^{-4}\Omega_{\rm p}$. When $1.3 < T_{\perp \alpha}/T_{\parallel \alpha} < 1.5$, the large majority of the data are constrained to lie below the curve corresponding to a maximum A/IC growth rate of $10^{-3} \Omega_{\rm p}$. Moreover, the curves corresponding to constant maximum growth rates have approximately the same slope as the contours of the probability distribution function (PDF) of the {\em Wind} data at $w_{\parallel \alpha} \gtrsim v_{\rm A}$.

\begin{figure}

\includegraphics[width=13pc,height=26pc]{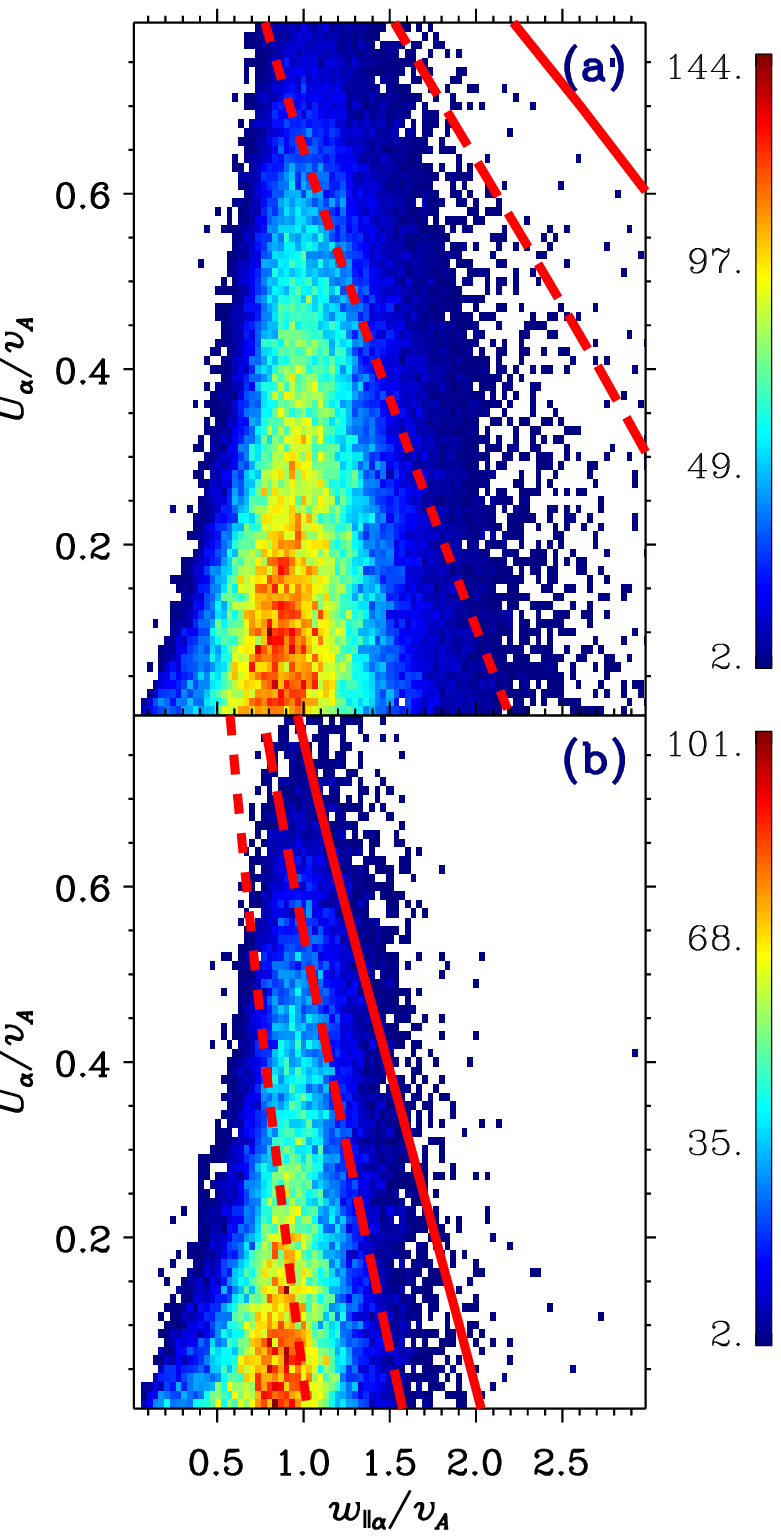}
\caption{ Distribution of data in the $U_\alpha - w_{\parallel
    \alpha}$ plane, where $U_\alpha$ is the alpha-proton drift speed
  and $w_{\parallel \alpha}$ is the parallel alpha-particle thermal
  speed. The number of measurements in each bin is shown in the color
  bars on the right.  The top panel is the subset of the data in which
  $0.7 < T_{\perp \alpha}/T_{\parallel \alpha} < 0.9$, and the bottom
  panel is the subset of the data in which $0.45 < T_{\perp
    \alpha}/T_{\parallel \alpha} < 0.55$.  The short-dashed lines are
  plots of the FM/W instability threshold ( Equation~(\ref{eq:Umin}))
  with $T_{\perp \alpha}/T_{\parallel \alpha} = 0.8$ (top panel) and
  $T_{\perp \alpha}/T_{\parallel \alpha} = 0.5$ (bottom panel).  The
  long-dash (solid) line correpsonds to parameter combinations for
  which the maximum FM/W growth rate is $\gamma = 10^{-3}\Omega_{\rm p}$
  ($\gamma =3\times 10^{-3}\Omega_{\rm p}$), where again $T_{\perp
    \alpha}/T_{\parallel \alpha} = 0.8$ in the top panel and $T_{\perp
    \alpha}/T_{\parallel \alpha} = 0.5$ in the bottom
  panel.} \label{fig.2}
\end{figure}

In Figure~\ref{fig.2}, we plot the instability threshold of the FM/W
wave from Equation~(\ref{eq:UFMW}) along with the subsets of the {\em Wind} data in which $0.7 <
T_{\perp \alpha}/T_{\parallel \alpha} < 0.9$ (top panel) and $0.45 <
T_{\perp \alpha}/T_{\parallel \alpha} < 0.55$ (bottom panel).  We also
plot curves corresponding to maximum FM/W growth rates of $10^{-3}
\Omega_{\rm p}$ and $3\times 10^{-3} \Omega_{\rm p}$.  When $0.7 <
T_{\perp \alpha}/T_{\parallel \alpha} < 0.9$, the majority of the data
have values of $U_\alpha$ smaller than the threshold value $U_{\rm
  FM/W}$ in Equation~(\ref{eq:UFMW}).  When $0.45 <
T_{\perp \alpha}/T_{\parallel \alpha} < 0.55$, a small fraction
of the data satisfies $U_\alpha > U_{\rm FM/W}$, but the majority of
the data is constrained to lie below the curve corresponing to $\gamma
= 10^{-3} \Omega_{\rm p}$.  In addition, the curves of constant maximum growth rates
and the contours of the PDF at $w_{\parallel \alpha} \gtrsim v_{\rm A}$ have similar slopes.

\begin{figure}
\includegraphics[width=20pc,height=20pc]{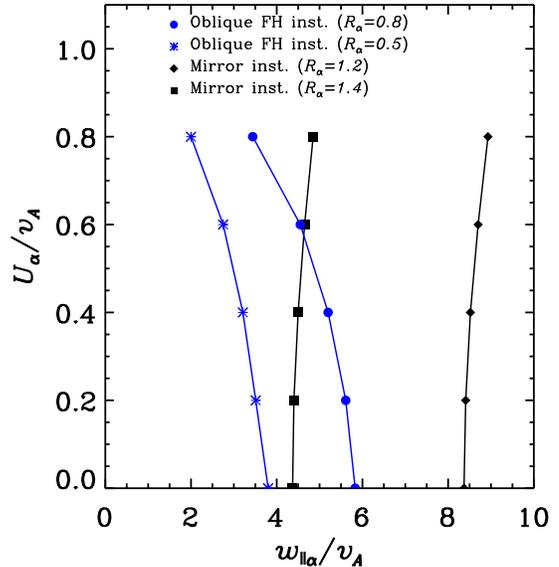}
\caption{ Isocontours of constant maximum growth rate $\gamma$ for the mirror-mode instability with $\gamma
= 10^{-3} \Omega_{\rm p}$ when $R_\alpha = 1.2$  (diamonds) and $R_\alpha = 1.4$ (squares).
Isocontours of constant maximum growth rate $\gamma$ for the oblique FH mode with $\gamma
= 10^{-3} \Omega_{\rm p}$, when $R_\alpha = 0.8$  (filled dots) and $R_\alpha = 0.5$ (asterisks).} \label{fig.3}
\end{figure}

We note that the constant-$\gamma$ contours for the parallel A/IC and
FM/W instability thresholds do not coincide with the contours of the
data distribution at small $w_{\parallel \alpha}/v_{\rm A}$ in
Figures~\ref{fig.1} and~\ref{fig.2}, where the upper bound on
$U_\alpha$ is approximately proportional to $w_{\parallel\alpha}$.
The reason for this upper bound on~$U_\alpha$ at
small~$w_{\parallel \alpha}/v_{\rm A}$ is not clear from our analysis.

Two other instabilities driven by pressure anisotropies are the mirror-mode and the oblique FH instabilities \citep{stix92}. If the temperature anisotropy crosses the instability threshold of the mirror-mode or the oblique FH instability, the unstable mode shows maximum growth rate at a non-vanishing  angle between the wavevector ${\bf k}$ and the background magnetic field ${\bf B}_0$. The frequencies of these oblique instabilities in the proton frame are purely imaginary if $U_{\alpha}=0$, and the real parts of the frequencies slowly increase with increasing $U_{\alpha}$. In Figure~\ref{fig.3} we plot numerically determined isocontours of constant maximum growth rates $\gamma$ for both the mirror-mode instability and the oblique FH instability in the $w_\parallel/v_A$-$U_\alpha/v_A $ plane for two different values of $R_\alpha$. The points represent parameter combinations for which the particular mode has $\gamma = 10^{-3}\Omega_{\rm p}$ at one wavevector only and has lower $\gamma$ at all other wavevectors.
Both the analytical thresholds and the isocontours with $\gamma = 10^{-3}\Omega_{\rm p}$ for the AI/C and FM/W instabilities are much closer to the data distribution in parameter space than the isocontours  for the oblique instabilities (compare Figure~\ref{fig.3} to Figures~\ref{fig.1} and \ref{fig.2}). Furthermore, the threshold of the mirror-mode instability hardly depends on the value of $U_\alpha$, and the slopes of the lines in Figure~\ref{fig.3} are very different from the slopes of the outer contours of the data distribution plotted in Figures~\ref{fig.1} and \ref{fig.2}. Therefore, we conclude that the oblique instabilities seem not to limit the alpha temperature anisotropy in the presence of alpha drift in our cases.

\begin{figure}
\includegraphics[width=13pc,height=19pc]{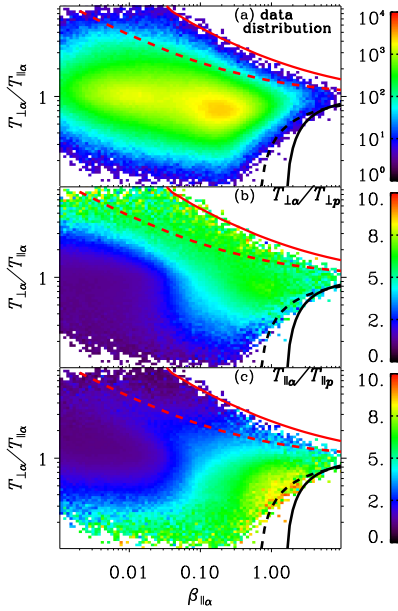}
\caption{ (a) Number of spectra in each bin. Average values of (b) $T_{\perp \alpha}/T_{\perp \rm p}$, 
and (c)  $T_{\parallel\alpha}/T_{\parallel \rm p}$
  are given by the color bars on the right. The upper red solid line (red dashed line)
represents parameter values for which the maximum  growth rate of the mirror-mode (A/IC) instability is $\gamma =  10 ^{-2} \Omega_{\rm p}$.  The lower black solid line (black dashed line) represents parameter values for which the maximum growth rate of the oblique FH (FM/W)  instability  is $\gamma = 10 ^{-2}\Omega_{\rm p}$.} \label{fig.4}
\end{figure}

We now turn to a consideration of the preferential heating of alpha
particles near the thresholds of the A/IC and FM/W instabilities. For this part of our analysis, we select all data points from the full {\em Wind} data set between 1994 and mid-2010 for which $U_\alpha<0.1 v_{\rm A}$.

In Figure~\ref{fig.4} we order the data as a function of $\beta_{\parallel
  \alpha}$ and $ T_{\perp\alpha}/T_{\parallel\alpha}$ (where $\beta_{\parallel \alpha}=2n_\alpha k_B T_{\parallel\alpha}\mu_0/B_0^2$). The curves in Figure~\ref{fig.4} are contours of constant maximum growth rate
$\gamma= 10^{-2} \Omega_{\rm p}$ using the analytical fitting formula (8) 
of \cite{maruca12} assuming isotropic proton temperature, $n_\alpha =0.05 n_{\rm p}$ and equal parallel thermal speeds of alpha particles and protons. The curves we plot thus serve primarily to indicate the vicinity of the
many different growth-rate contours that would apply to this data set.
In the top panel of Figure~\ref{fig.4}, we plot the data distribution
as a function of $\beta_{\parallel\alpha}$ and $R_\alpha$. In the panels (b) and (c) of Figure~\ref{fig.4}, we plot
the average value of $T_{\perp \alpha}/T_{\perp p}$ and $T_{\parallel
  \alpha}/T_{\parallel p}$, respectively. These plots show that the ratio $T_{\perp \alpha}/T_{\perp p}$ ($T_{\parallel\alpha}/T_{\parallel p}$) is relatively higher near the threshold of the A/IC (FM/W) instability than elsewhere in the ($\beta_{\parallel
  \alpha}$--$ T_{\perp\alpha}/T_{\parallel\alpha}$) plane. This observational finding for alpha particles, to the
best of our knowledge, has not been reported before. However,
\citet{maruca11} reported a similar finding for protons.  Although we
do not focus on the origin of the enhanced alpha-particle temperatures
in this study, we note that cyclotron-heating and stochastic-heating models
can explain the preferential heating of alpha particles to temperatures exceeding the proton temperature~\citep{isenberg07,kasper13,chandran13}.

\vspace{0.2cm} 
\section{Conclusions}
\vspace{0.2cm}

By analyzing {\em Wind} measurements of solar-wind streams, we find that  the alpha-particle
differential flow is limited to values comparable to the instability
thresholds of A/IC and FM/W waves. Importantly, these thresholds
depend upon the temperature anisotropy of the alpha particles.  In
contrast to the $U_{\alpha}$ thresholds of beam instabilities in
isotropic-temperature plasmas, which are~$\gtrsim v_{\rm A}$, the
thresholds of the A/IC and FM/W instabilities can be significantly
smaller than~$v_{\rm A}$ when $T_{\perp \alpha} \neq T_{\parallel
  \alpha}$ and when $w_{\parallel \alpha} \gtrsim v_{\rm A}$.  Our
findings support previous suggestions that A/IC and FM/W instabilities
limit the alpha-particle differential flow in the solar wind. Our results
also emphasize the importance of treating differential flow and
temperature anisotropy on an equal footing when $w_{\parallel \alpha}
\gtrsim v_{\rm A}$, since these properties are of comparable
importance for these instabilities. 

Within the subset of the data in which~$U_{\alpha} < 0.1 v_{\rm A}$,
we find strong preferential heating of alpha particles relative to
protons for conditions under which the A/IC and FM/W instabilities
occur. When the plasma is near the threshold of the A/IC instability,
$T_{\perp \alpha}/T_{\perp \rm p}$ is unusually large.  On the other
hand, when the plasma is near the threshold of the FM/W instability,
$T_{\parallel \alpha}/T_{\parallel \rm p}$ is unusually large. This
suggests that exceptionally strong perpendicular (parallel) heating is
the reason why, in a small fraction of the small-$U_\alpha$ data,
alpha-particles are in the A/IC-unstable (FM/W-unstable) region of
parameter space.

\end{document}